\documentclass[useAMS,usenatbib,usegraphicx]{mn2e}

\title[Weighing dark matter haloes]{Weighing dark matter haloes with gravitationally lensed supernovae}
\author[J. J\"onsson et al.]{
J. J\"onsson$^{1}$\thanks{E-mail:jacke@astro.ox.ac.uk}, 
T. Dahl\'en$^{2}$ ,
I. Hook$^{1,3}$,
A. Goobar$^{4,5}$, and
E. M\"ortsell$^{4,5}$ \\
$^{1}$University of Oxford Astrophysics, Denys Wilkinson
  Building, Keble Road, Oxford OX1 3RH, UK \\
$^{2}$Space Telescope Science Institute, 3700 San Martin Drive, Baltimore, MD 21218, USA \\
$^{3}$INAF - Osservatorio Astronomico di Roma, via Frascati 33, 00040
Monteporzio (RM), Italy \\
$^{4}$Physics Department, Stockholm University, AlbaNova University Center, SE--106 91,
Stockholm, Sweden \\
$^{5}$The Oskar Klein Center, Stockholm University, SE--106 91,
Stockholm, Sweden}
\begin{document}
%\date{Accepted . Received; in original form}

%\pagerange{\pageref{firstpage}--\pageref{lastpage}} \pubyear{XXXX}

\maketitle

\label{firstpage}

%%%%%%%%%%%%%%%%%%%%%%%%%%%%%%%%%%%%%%%%%%%%%%%%%%%%%%%%%%%%%%%%%%%%%%%%%%%%%
%%%%%%%%%%%%%%%%%%%%%%%%%%%%%%%%%%%%%%%%%%%%%%%%%%%%%%%%%%%%%%%%%%%%%%%%%%%%%
\begin{abstract}
High redshift Type Ia supernovae (SNe~Ia) are likely to be gravitationally lensed by dark matter haloes of galaxies in the foreground. Since SNe~Ia have very small dispersion after light curve shape and colour corrections, their brightness can be used to measure properties of the dark matter haloes via gravitational magnification.
We use observations of galaxies and SNe~Ia within the Great Observatories Origins Deep Survey (GOODS) to measure the relation between
galaxy luminosity and dark matter halo mass. 
The relation we investigate is a scaling law between velocity dispersion and galaxy luminosity in the 
$B$-band: $\sigma=\sigma_*(L/L_*)^{\eta}$, where $L_*=10^{10}h^{-2}L_{\sun}$. 
The best-fitting values to this relation are
$\sigma_*=136$ km s$^{-1}$ and $\eta=0.27$. 
We find $\sigma_* \la 190$ km s$^{-1}$ at the 95 per cent confidence level. 
This method provides an independent cross-check of measurements of dark matter halo properties from galaxy--galaxy lensing studies. 
Our results agree with the galaxy--galaxy lensing results, but have much larger uncertainties. 
The GOODS sample of SNe Ia is relatively small (we include 24 SNe) and the
results therefore depend on individual SNe~Ia.
We have investigated a number of potential systematic effects. Light curve fitting, which affects the inferred brightness of the SNe~Ia, appears to be the most important one. 
Results obtained using different light curve fitting procedures differ at the 68.3 per cent confidence level.
\end{abstract}


%%%%%%%%%%%%%%%%%%%%%%%%%%%%%%%%%%%%%%%%%%%%%%%%%%%%%%%%%%%%%%%%%%%%%%%%%%%%%
%%%%%%%%%%%%%%%%%%%%%%%%%%%%%%%%%%%%%%%%%%%%%%%%%%%%%%%%%%%%%%%%%%%%%%%%%%%%%
\section{Introduction}
Weak gravitational lensing 
can lead to stretched or magnified (or de-magnified) images of distant sources. 
The stretching of an image depends on the shear of the lens, whereas the magnification 
is related to the convergence of the lens.
Measurements of shear through the ellipticity of galaxies lensed by galaxies 
(galaxy--galaxy lensing) have successfully been used to probe the nature of dark matter haloes 
\citep{tys84,bra96,hud98,fis00,guz02,hoe03,hoe04,kle06}. 
In this paper we use measurements of 
convergence rather than shear to investigate properties of 
dark matter haloes. 

The convergence can be obtained from the brightness of standard candles, because magnified standard candles 
are brighter than de-magnified ones.  As standard candles we use SNe~Ia,
which after light curve shape and colour corrections have small dispersion. 
Gravitational lensing adds extra scatter to the Hubble diagram \citep{kan95,fri97,wam97,hol98} and is
consequently regarded as a systematic uncertainty in the context of SN~Ia cosmology.
This extra dispersion, which increases with redshift \citep[see, e.g.,][]{ber00,hol05,gun06}, can also be used as a cosmological probe \citep{met99,dod06}. 
The distribution of SN~Ia Hubble diagram residuals, which is assumed to trace the magnification probability distribution, 
can be used as another gravitational lensing probe \citep{wan05}, 
which is sensitive to the fraction of compact objects in the Universe \citep{rau91,metsil99,sel99,mor01,min02,met07}.

Here we use yet another method which utilises individual SNe~Ia rather than the distribution of residuals.
We cross-correlate SN~Ia brightnesses
and gravitational magnifications computed using properties of galaxies in the foreground, as suggested by \citet{met98}. 
For an alternative, potentially very powerful method, also utilising correlations between foreground galaxies and
supernova brightnesses, see \citet{met01}.  

For our investigation we use a sample of SNe~Ia which was obtained within the Great Observatories Origins Deep Survey (GOODS).
Owing to the {\it HST} observations, which were part of the GOODS, the sample contain some of the most distant SNe~Ia ever 
observed \citep{rie04,rie07}. The high redshift of the supernovae means that they are likely to be significantly magnified 
by the matter in the foreground \citep{ber00,hol05}. This magnification is then used to infer the properties of the lensing galaxies.

The outline of the paper is as follows. 
In section~\ref{sec:data} we describe the sample of SNe~Ia and the galaxy catalogues we use. The method is explained in section~\ref{sec:method}.
The results obtained when applying the method to the data are presented in section~\ref{sec:sisresult}.
Systematic effects which could affect the results are investigated in section~\ref{sec:sys}. Section~\ref{sec:union} is devoted 
to a particularly important systematic effect, namely  
the differences in the results which arise due to different light curve fitting packages.
The results are finally summarised and discussed in section~\ref{sec:disc}.

Throughout the paper, we assume a flat universe dominated by dark matter and a cosmological constant. Unless otherwise stated 
$\Omega_{\rm M}=0.3$ and $\Omega_{\Lambda}=1-\Omega_{\rm M}$.
 
%%%%%%%%%%%%%%%%%%%%%%%%%%%%%%%%%%%%%%%%%%%%%%%%%%%%%%%%%%%%%%%%%%%%%%%%%%%%%
%%%%%%%%%%%%%%%%%%%%%%%%%%%%%%%%%%%%%%%%%%%%%%%%%%%%%%%%%%%%%%%%%%%%%%%%%%%%%
\section{Data} \label{sec:data}
In this paper we make use of data observed within the GOODS project. Two fields were observed within this project:
GOODS--North ($12^{\rmn{h}}~36^{\rmn{m}}~55^{\rmn{s}}$, $+62\degr~14\arcmin~15\arcsec$) 
and GOODS-South ($3^{\rmn{h}}~32^{\rmn{m}}~30^{\rmn{s}}$,$-27\degr~48\arcmin~20\arcsec$). The observations include both galaxies and supernovae.

\subsection{Supernovae}
Thanks to the {\it HST} observations, which formed part of the GOODS project,
some of the furthest supernovae ever discovered were located in GOODS--S and GOODS--N  \citep{rie04,str04,rie07}.
We also include SN1997ff \citep{rie01} at $z \simeq 1.8$  located in GOODS--N in our sample, although it was not detected in the actual GOODS search.
Since the effects of gravitational lensing increase with redshift, the GOODS SNe~Ia are well suited to our purposes. 
For each SN~Ia we use redshift, distance modulus corrected for dust extinction, and its uncertainty from \citet{rie07}. 
The SNe~Ia in \citet{rie07} are classified as either 
gold (high confidence) or silver (likely but uncertain). We work primarily with the gold SNe~Ia. 
There are 7 and 19 gold SNe~Ia located in the GOODS--S and GOODS--N, respectively.  
Our Gold sample hence consists of 26 SNe~Ia.
In order to compute the magnification,
we need the redshift, $z_{\rm SN}$, 
and the position on the sky, $\btheta_{\rm SN}$, for all SNe~Ia. We will refer to these variables as
\begin{equation}
\blambda_{\rm SN} =\{z_{\rm SN},\btheta_{\rm SN}\},
\end{equation}
in the following.

\subsubsection{Hubble diagram residuals}
In order to compute the Hubble diagram residuals, $\Delta m_{\rm SN}$, which tell us whether a SN~Ia is brighter or dimmer than average, we
have to predict the average SN~Ia brightness (i.e.~distance modulus)  as a function of redshift. To predict the distance modulus
we use the formula
\begin{equation}
\mu_{\rm p}(z)=\mathcal{M}+5\log_{10}D_{\rm L}(z;\Omega_{\rm M}),
\end{equation}
where the $H_0$-independent luminosity distance,
\begin{equation}
D_{\rm L}(z;\Omega_{\rm M})=(1+z)\int_0^z \frac{dz'}{\sqrt{\Omega_{\rm M}(1+z')^3+(1-\Omega_{\rm M})}},
\end{equation}
depends only on $\Omega_{\rm M}$, since we assume the Universe to be flat, and
the redshift.
We use the best-fitting value of the normalisation constant  $\mathcal{M}$ to the Gold sample \citep{rie07} 
obtained for a fixed value of $\Omega_{\rm M}$. 
Including also the SNe~Ia classified as silver in \citet{rie07} makes a negligible difference to $\mathcal{M}$ of -0.008 mag.
The Hubble diagram 
residual of a SN~Ia is thus computed by
\begin{equation}
\Delta m_{\rm SN}=\mu_0-\mu_{\rm p},
\end{equation} 
where $\mu_0$ is the extinction corrected distance modulus of the supernova from \citet{rie07}. 
We assume the uncertainty in the Hubble diagram residual, $\sigma_{\Delta m_{\rm SN}}$, to be the 
same as the uncertainty in the distance modulus \citep{rie07}. 
The supernova data and the residuals computed for $\Omega_{\rm M}=0.3$ are listed in Table~\ref{tab:sn}. 

\subsubsection{Field boundaries} \label{sec:bound}
Since unobserved galaxies outside of the fields can contribute to the magnification, only SNe~Ia located 
further than a distance $\theta_{\rm c}$ on the sky from the boundaries of the fields are included in the analysis. In \citet{gun06}
it was shown that it is sufficient to include galaxies within a circular radius of $60\arcsec$ when estimating the 
magnification of SNe~Ia due to foreground galaxies. We therefore use $\theta_{\rm c}=60\arcsec$ as our cut-off radius.
HST04Sas is located too close to the boundary of GOODS--N and is therefore excluded from the Gold sample.
HST05Koe is located in a masked region, which lack observations of foreground galaxies, and is for that reason also excluded from the Gold sample.
We consequently use only 24 of the SNe~Ia in the Gold sample.

\subsection{Galaxies}
To model the dark matter haloes in the foreground of the SNe~Ia we use galaxy observations obtained within the GOODS
\citep{gia04,cap04}. 
For GOODS-S we use a VLT/ISAAC selected catalog with a 10$\sigma$~point source sensitivity of 
$m_{\rm AB}=25.1$ mag \citep{gia04}, while we for GOODS-N use a SUBARU/Supreme cam $R$-band selection with a 5$\sigma$~point source sensitivity of $m_{\rm AB}=26.6$ mag \citet{cap04}.
The GOODS--N and the GOODS--S have been imaged in many wavelength bands, ranging from the $U$-band to the $K$-band,
 which make it possible to calculate accurate photometric redshifts for the galaxies in these fields.

\subsubsection{Photometric redshifts}
Since our galaxy catalogues contain several thousands of objects, most of them have only photometric redshifts. To compute the photometric redshifts we
use a version of the template-fitting method \citep{dah05}. This method gives us for each galaxy
a photometric redshift distribution, an absolute $B$-band magnitude, and the best-fitting spectral type. The probability distribution has sometimes multiple peaks, reflecting different possible solutions. Depending on the redshift of the SN~Ia, different peaks could place the galaxy in the foreground or background of the supernova. The photometric redshift distributions,
$p(z)$,
 are in most cases asymmetric and the mode, i.e.~the highest peak, is usually different from 
the average redshift. 
As our photometric redshift estimate for the galaxies in the line of sight we generate (for each one)
a random number from the derived probability distribution. \citet{wit09} has shown that this is a
more robust approach than using the mean or median of the distribution.

In order to compute the magnification of a SN~Ia we need properties of the $N_{\rm gal}$ galaxies along the line of sight. 
These properties will in the following be referred to as
\begin{equation}
\blambda_{\rm los}=\{z_{\rm gal}^1,\btheta_{\rm gal}^1,M_B^1,\tau^1, \ldots, 
z_{\rm gal}^{N_{\rm gal}},\btheta_{\rm gal}^{N_{\rm gal}},
M_B^{N_{\rm gal}},\tau^{N_{\rm gal}}\},
\end{equation} 
where $z_{\rm gal}^i$, $\btheta_{\rm gal}^i$,  $M_B^i$, and $\tau^i$ are the redshift, position on the sky,
absolute $B$-band magnitude, and spectral type of the $i$th galaxy, respectively.

% SN data table
%%%%%%%%%%%%%%%%%%%%%%%%%%%%%%%%%%%%%%%%%%%%%%%%%%%%%%%%%%%%%%%%%%%%%%%%%%%%%
 %{@{}lcccccc}
\begin{table*}
\centering
\begin{minipage}{140mm}
 \renewcommand{\thefootnote}{\alph{footnote}} 
 \caption{GOODS SN~Ia data.}
\label{tab:sn}
 \begin{tabular}{lllllr@{.}ll}
  \hline
  Name & Sample\footnotemark[1] & $z_{\rm SN}$ & $\alpha_{\rm SN}$ (J2000)  & $\delta_{\rm SN}$ (J2000)
        & \multicolumn{2}{c}{$\Delta m_{\rm SN}$ (mag)} & $\sigma_{\Delta m_{\rm SN}}$ (mag) \\
  \hline
  HST04Kur  & Silver &0.36 &  $03^{\rmn{h}}~32^{\rmn{m}}~36\fs03$& $-27\degr~51\arcmin~17\farcs66$ & $-$0&40 &  0.39 \\
  HST04Yow & Gold, Union & 0.457 & $12^{\rmn{h}}~36^{\rmn{m}}~34\fs33$ & $+62\degr~12\arcmin~12\farcs95$  & $-$0&01 & 0.32 \\
  HST04Haw\footnotemark[2] 
    & Silver &0.49 & $12^{\rmn{h}}~35^{\rmn{m}}~41\fs16$ &  $+62\degr~11\arcmin~37\farcs19$  & 0&11 &  0.24 \\
  SN2002hr  & Silver, Union & 0.53 & $03^{\rmn{h}}~32^{\rmn{m}}~22\fs57$  & $-27\degr~41\arcmin~52\farcs20$ &0&47 & 0.27\\
  HST05Zwi & Silver, Union &  0.52 & $03^{\rmn{h}}~32^{\rmn{m}}~45\fs65$ &  $-27\degr~44\arcmin~ 24\farcs30$   &$-$0&54 &  0.37 \\
 SN2003be  & Gold&  0.64 &  $12^{\rmn{h}}~36^{\rmn{m}}~25\fs97$ & $+62\degr~06\arcmin~55\farcs60$  & $-$0&10 & 0.25 \\
 HST05Dic\footnotemark[2]
   & Silver & 0.64 &  $12^{\rmn{h}}~35^{\rmn{m}}~49\fs61$ & $+62\degr~10\arcmin~11\farcs96$  &$-$0&23 &0.18\\
SN2003bd  & Gold&  0.67 &  $12^{\rmn{h}}~37^{\rmn{m}}~25\fs06$ & $+62\degr~13\arcmin~17\farcs50$ &$-$0&05 & 0.24 \\
HST04Rak & Gold, Union& 0.739   &$ 03^{\rmn{h}}~32^{\rmn{m}}~18\fs15$ & $-27\degr~44\arcmin~10\farcs55$  &  $-$0&12 & 0.22 \\
SN2002kd  & Gold, Union&  0.74  &  $03^{\rmn{h}}~32^{\rmn{m}}~22\fs34$ & $-27\degr~44\arcmin~26\farcs90$   & $-$0&34  & 0.19 \\
HST05Spo  & Gold, Union& 0.839  & $12^{\rmn{h}}~37^{\rmn{m}}~06\fs53$ & $+62\degr~15\arcmin~11\farcs70$ &  $-$0&39 & 0.20 \\
SN2003eq & Gold, Union&   0.85 &  $12^{\rmn{h}}~37^{\rmn{m}}~48\fs34$ & $+62\degr~13\arcmin~35\farcs30$ & $-$0&17  & 0.21 \\
HST04Man & Gold, Union& 0.854 & $12^{\rmn{h}}~36^{\rmn{m}}~34\fs81$& $+62\degr~15\arcmin~49\farcs06$ & 0&07 & 0.29 \\
SN2003eb  & Gold&  0.90  & $12^{\rmn{h}}~37^{\rmn{m}}~15\fs18$ &  $+62\degr~13\arcmin~34\farcs60$ & $-$0&39 & 0.25 \\
SN2003es   & Gold& 0.954 & $12^{\rmn{h}}~36^{\rmn{m}}~55\fs39$ & $+62\degr~13\arcmin~11\farcs90 $ & 0&12 & 0.27 \\
HST04Tha  & Gold& 0.954  &$12^{\rmn{h}}~36^{\rmn{m}}~55\fs17$ & $+62\degr~13\arcmin~04\farcs05$  & $-$0&33 & 0.27 \\
HST04Omb & Gold, Union& 0.975 &  $03^{\rmn{h}}~32^{\rmn{m}}~25\fs34$ & $-27\degr~45\arcmin~03\farcs01$  &  $-$0&03  & 0.26 \\
HST04Eag  & Gold, Union& 1.019  & $12^{\rmn{h}}~37^{\rmn{m}}~20\fs75$ & $+62\degr~13\arcmin~41\farcs50$ &  0&16 &  0.19 \\
HST05Fer  & Gold, Union& 1.020  & $12^{\rmn{h}}~36^{\rmn{m}}~25\fs10$ & $+62\degr~15\arcmin~23\farcs84$ & $-$0&37 & 0.27 \\
HST05Str  & Gold, Union& 1.027  & $12^{\rmn{h}}~36^{\rmn{m}}~20\fs63$ & $+62\degr~10\arcmin~50\farcs58$  & 0&43 & 0.19 \\
HST05Gab & Gold, Union& 1.12 &  $12^{\rmn{h}}~36^{\rmn{m}}~13\fs83$& $+62\degr~12\arcmin~07\farcs56$  &  0&05 & 0.29 \\
SN2002ki  & Gold, Union&  1.14  & $12^{\rmn{h}}~37^{\rmn{m}}~28\fs35$ & $+62\degr~20\arcmin~40\farcs00$  &  0&05 & 0.29 \\
HST04Gre & Gold, Union& 1.14 &   $03^{\rmn{h}}~32^{\rmn{m}}~21\fs49$ & $-27\degr~46\arcmin~58\farcs30$ & $-$0&22 & 0.31 \\
 HST05Red & Silver, Union & 1.19 & $12^{\rmn{h}}~37^{\rmn{m}}~01\fs70$ & $+62\degr~12\arcmin~23\farcs98$  & $-$1&14 & 0.39 \\
HST05Koe\footnotemark[2]
 & Gold& 1.23 & $12^{\rmn{h}}~36^{\rmn{m}}~22\fs92$ & $+62\degr~18\arcmin~23\farcs20$ & 0&30   &0.23 \\
HST05Lan  & Gold, Union&1.235 & $12^{\rmn{h}}~36^{\rmn{m}}~56\fs72$ & $+62\degr~12\arcmin~53\farcs33$ &  0&11 & 0.20 \\
SN2003az  & Silver, Union & 1.26 & $12^{\rmn{h}}~37^{\rmn{m}}~19\fs67$ & $+62\degr~18\arcmin~37\farcs50$  & $-$0&31 &  0.25 \\
SN2002fw  & Gold, Union&  1.30 &  $03^{\rmn{h}}~32^{\rmn{m}}~37\fs52$ & $-27\degr~46\arcmin~46\farcs60$  &   0&04 & 0.20 \\
SN2002hp  & Gold, Union&  1.30 &  $03^{\rmn{h}}~32^{\rmn{m}}~24\fs79$ & $-27\degr~46\arcmin~17\farcs80$  &   $-$0&52 & 0.30 \\
SN2003aj\footnotemark[2]
  & Silver & 1.31 &  $03^{\rmn{h}}~32^{\rmn{m}}~44\fs33$ & $-27\degr~55\arcmin~06\farcs40$ & $-$0&05 & 0.31 \\
SN2003dy  & Gold, Union&  1.34  & $12^{\rmn{h}}~37^{\rmn{m}}~09\fs16$ &  $+62\degr~11\arcmin~29\farcs00$  & $-$0&17  & 0.31 \\
HST04Mcg & Gold, Union& 1.357 & $03^{\rmn{h}}~32^{\rmn{m}}~10\fs02$ & $-27\degr~49\arcmin~49\farcs98$   &  0&07 & 0.25 \\
HST04Sas\footnotemark[3]
 & Gold, Union& 1.39  & $12^{\rmn{h}}~36^{\rmn{m}}~54\fs11$ & $+62\degr~08\arcmin~22\farcs76$  & $-$0&29 & 0.19 \\
 SN2002fx  & Silver & 1.40 &  $03^{\rmn{h}}~32^{\rmn{m}}~06\fs80$ & $-27\degr~44\arcmin~34\farcs40$ & 0&06 &  0.81 \\
 SN2003ak\footnotemark[2]
  & Silver, Union & 1.55 & $03^{\rmn{h}}~32^{\rmn{m}}~46\fs90$ &  $-27\degr~54\arcmin~49\farcs30$ & $-$0&43 &  0.32 \\
SN1997ff  & Gold&  1.755  & $12^{\rmn{h}}~36^{\rmn{m}}~44\fs10$ & $+62\degr~12\arcmin~44\farcs69$   &  $-$0&48 & 0.35 \\
  \hline
 \end{tabular}
  \footnotetext[1]{The terms Gold and Silver indicate that the SN~Ia has been classified as gold or silver by 
  \citet{rie07}. The term Union indicates that the SN~Ia belongs to the compilation of \citet{kow08}}
 \footnotetext[2]{Located too close to the boundary of the field.}
  \footnotetext[3]{Located too close to a masked region.}
\end{minipage}
\end{table*}
  
%%%%%%%%%%%%%%%%%%%%%%%%%%%%%%%%%%%%%%%%%%%%%%%%%%%%%%%%%%%%%%%%%%%%%%%%%%%%%
%%%%%%%%%%%%%%%%%%%%%%%%%%%%%%%%%%%%%%%%%%%%%%%%%%%%%%%%%%%%%%%%%%%%%%%%%%%%%  
\section{Method} \label{sec:method}

\subsection{Computing the magnification}
When computing the magnification of an individual SN~Ia, we take 
many galaxies along the line of sight into account. 
Since the thickness of the lenses is small compared to the distances between them, we can use the
thin screen approximation. In this approximation the lenses are treated as two-dimensional surfaces 
perpendicular to the line joining the source and the observer. The surface mass density of a lens, $\Sigma(\bxi)$,
is obtained by projecting the mass density, $\rho(\bmath{r})$, of the lens onto a plane (the lens plane),
\begin{equation}
\Sigma(\bxi)=\int_{-\infty}^{\infty} \rho(\bxi,y)dy,
\label{eq:proj}
\end{equation} 
where $\bxi$ is a vector in the lens plane and $y$ is a coordinate along the line of sight. In this step
of the calculations, the density profile of the dark matter haloes enters. 

In the weak lensing approximation
the magnification factor, $\mu$, is obtained by summing the convergence due to all lenses along the line of sight, 
\begin{equation}
\mu \simeq 1+2\sum_{i=1}^{N_{\rm gal}} \kappa^i_{\rm gal}.
\label{eq:mu}
\end{equation}
The convergence of a single galaxy halo,
\begin{equation}
\kappa_{\rm gal}=\Sigma/\Sigma_{\rm c},
\label{eq:kapdef}
\end{equation}
is the surface density scaled by the critical surface density, 
$\Sigma_{\rm c}=D_{\rm s}/(4\pi G D_i D_{i{\rm s}})$,
where $D_{\rm s}$ is the angular diameter distance between the observer and the source, 
$D_i$ is the distance between the observer and the $i$th lens, and $D_{i{\rm s}}$ is the 
distance between the $i$th lens and the source.

For most SNe~Ia we expect the effects of gravitational lensing to be small. In the case of small magnifications the weak lensing approximation is appropriate. Since our sample contains SNe~Ia at very high redshift the weak lensing approximation might not be valid for all SNe~Ia. We therefore 
recursively trace a light ray through all lens planes to find the position of the SN~Ia in the source plane and its magnification.

The position of the SN~Ia in the $j$th lens plane can be found from
\begin{equation}
\bxi_j=\frac{D_j}{D_1}\bxi_1-\sum_{i=1}^{j-1}D_{ij}\hat{\balpha}_{\rm gal}^i(\bxi_i),
\label{eq:trace}
\end{equation}
where $\hat{\balpha}_{\rm gal}^i$ is the deflection angle due to the $i$th galaxy halo. 
$D_{ij}$ represents the distance between the $i$th and the $j$th lens plane.
For spherically symmetric lenses, the deflection angle, which depends on the surface mass density, is given by
\begin{equation}
\hat{\balpha}_{\rm gal}(\bxi)=\frac{4G}{\xi}2\pi\int_0^\xi \Sigma(\xi')\xi'd\xi' \hat{\bxi},
\end{equation}
where $\xi=|\bxi|$ and $\hat{\bxi}$ is a unit vector in the direction of $\bxi$.

The magnification matrix can also be computed recursively by
\begin{equation}
\mathbfss{A}_j=\mathbfss{I}-\sum_{i=1}^{j-1}\beta_{ij}\mathbfss{U}_i\mathbfss{A}_i,
\end{equation}
where $\mathbfss{I}$ is the unit matrix and $\mathbfss{A}_1=\mathbfss{I}$.
In the above equation $\beta_{ij}$
represents the distance combination $\beta_{ij}=D_{ij}D_{\rm s}/(D_jD_{i\rm s})$, where 
$D_{\rm s}$ and $D_{i\rm s}$ are the distances between the observer and the source and between the $i$th
lens plane and the source, respectively.
The matrix 
\begin{equation}
\mathbfss{U}=\left( \begin{array}{cc}
\kappa_{\rm gal}+\gamma_{{\rm gal},1} & \gamma_{{\rm gal},2} \\
\gamma_{{\rm gal},2} & \kappa_{\rm gal}-\gamma_{{\rm gal},1} 
\end{array} 
\right),
\end{equation}
depends on the convergence, $\kappa_{\rm gal}$, and the shear components,  
$\gamma_{{\rm gal},1}$ and $\gamma_{{\rm gal},2}$, of the lens.
The magnification factor of the SN~Ia is given by
\begin{equation}
\mu=\frac{1}{\det{\mathbfss{A}}_{N_{\rm gal}+1}},
\label{eq:muray}
\end{equation}
i.e.~ the inverse of the determinant of the magnification matrix corresponding to the source plane 
($j=N_{\rm gal}+1$).

In order to compute the magnification of a SN~Ia relative to a homogeneous universe, the magnification obtained via equation~(\ref{eq:muray}) needs to be normalised using the average magnification factor,
$\langle \mu \rangle$. 
In terms of logarithmic magnitudes the magnification of a SN~Ia is given by, 
\begin{equation}
\Delta m_{\rm lens}=-2.5\log_{10}\mu+2.5\log_{10}\langle \mu \rangle.
\end{equation}  
To compute $\langle \mu \rangle$ we use simulated lines of sight. 
For the simulated lines of sight we use the weak lensing approximation, i.e.~equation~(\ref{eq:mu}).
See section~\ref{sec:norm} for a discussion of this approach.

\subsection{Halo model}
To model dark matter haloes we use singular isothermal spheres (SISs). The SIS model is characterised by a single parameter:
the velocity dispersion, $\sigma$.
Since the SIS model is axially symmetric, the surface density, $\Sigma(\xi)=\sigma^2/(2G\xi)$, is a function of
the physical distance between the source and the lens in the source plane, $\xi$.
Given the distance
to the lens, $D_i$, and the angular separation between the source and the lens, $\theta$, the physical distance in the lens plane 
can be computed by 
the formula $\xi=D_i\theta$.
For a SIS the convergence is given by
\begin{equation}
\kappa_{\rm gal}(x)=\frac{1}{2x},
\end{equation}
where $x=\xi/\xi_0$ and 
\begin{equation}
\xi_0=4\pi  \sigma^2 \frac{D_i D_{i{\rm s}}}{D_{\rm s}}.
\label{eq:sisxi}
\end{equation}
The shear components of a SIS are
\begin{equation}
\gamma_{{\rm gal},1}(\mathbf{x})=\frac{x_2^2-x_1^2}{2x^3}
\end{equation} 
and 
\begin{equation}
\gamma_{{\rm gal},2}(\mathbf{x})=-\frac{x_1x_2}{x^3},
\end{equation}
where $x_1$ and $x_2$ are the two components of the vector $\mathbf{x}=\bxi/\xi_0$.

To investigate the relationship
between galaxy luminosity and velocity dispersion, we use the scaling law
\begin{equation}
\sigma=\sigma_{*}\left( \frac{L}{L_*} \right)^{\eta},
\label{eq:sislaw}
\end{equation}
which has two parameters: $\sigma_*$ and $\eta$.
$L_*$ is a characteristic luminosity, which we take to be $10^{10}h^{-2}L_{\sun}$ in the $B$-band. 
In terms of absolute $B$-band magnitudes, which we will work with, the scaling relation becomes
\begin{equation}
\sigma=\sigma_*10^{-\eta(M_B-M_B^*)/2.5}, 
\end{equation}
where $M_B^*=-19.52+5\log_{10} h$. Our estimates of $M_B$ for the galaxies depends on $h$ in the same way as $M_B^*$,
which means that $h$ (the Hubble constant is related to this parameter through $H_0=100h$ km s$^{-1}$ Mpc$^{-1}$)
 cancel out from the calculations. For GOODS--S and GOODS--N, the average absolute $B$-band magnitude of the galaxies 
are  $\langle M_B \rangle=-19.46+5\log_{10} h$ and $\langle M_B \rangle=-19.13+5\log_{10} h$, respectively. The average 
values of $M_B$ are thus close to the value of $M_B^*$.

Our model describes a scaling relation between luminosity and velocity dispersion with two parameters, which we will refer to as
\begin{equation}
\blambda_{\rm halo}=\{\sigma_*,\eta\}.
\end{equation}

\subsection{$\chi^2$-statistic}
In order to find the best-fitting halo model, with parameters $\blambda_{\rm halo}$, we minimise
the following $\chi^2$-statistic:  
\begin{equation}
\chi^2=\sum_{i=1}^{N_{\rm SN}} \left[ \frac{
\Delta m_{\rm SN}^i-\Delta m_{\rm lens}(\blambda_{\rm SN}^i,\blambda_{\rm los}^i;\blambda_{\rm halo})}
{\sigma^i_{\Delta m_{\rm SN}}}
 \right]^2,
\label{eq:chi2} 
\end{equation}
where the superscripts refer to the $i$th SN~Ia,
The predicted magnification of a SN~Ia, 
$\Delta m_{\rm lens}(\blambda_{\rm SN},\blambda_{\rm los};\blambda_{\rm halo})$, 
depends on the halo model $\blambda_{\rm halo}$ and the variables $\blambda_{\rm SN}$ describing the SN~Ia position
and the variables $\blambda_{\rm los}$ describing the galaxies along the line of sight.

%%%%%%%%%%%%%%%%%%%%%%%%%%%%%%%%%%%%%%%%%%%%%%%%%%%%%%%%%%%%%%%%%%%%%%%%%%%%%
%%%%%%%%%%%%%%%%%%%%%%%%%%%%%%%%%%%%%%%%%%%%%%%%%%%%%%%%%%%%%%%%%%%%%%%%%%%%%
\section{Results} \label{sec:sisresult}
\begin{figure}
\includegraphics[width=84mm,angle=-90]{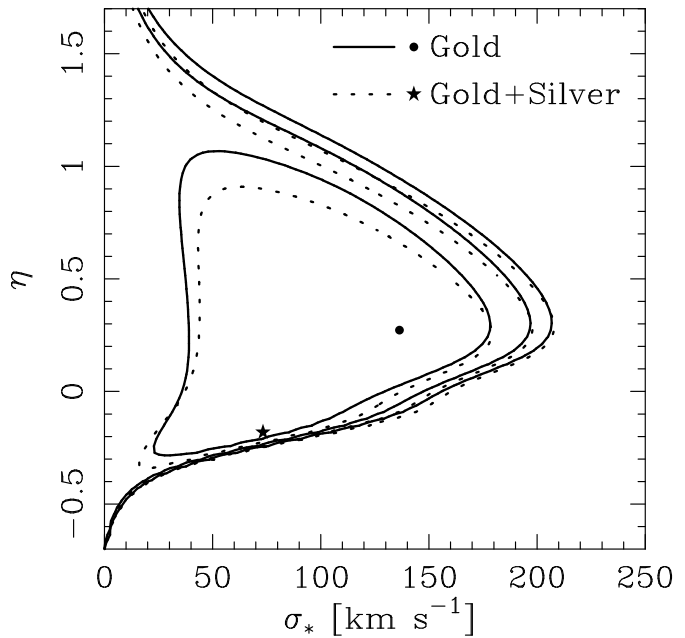}   
 \caption{
 Confidence level contours (68.3, 95, and 99 per cent) 
 in the $(\sigma_*,\eta)$-plane. 
 Circle and solid contours indicate best-fitting values
($\sigma_*=136$ km s$^{-1}$ and $\eta=0.27$)
and confidence level contours obtained for the Gold sample.
Star and dotted contours correspond to best-fitting values 
($\sigma_*=73$ km s$^{-1}$ and $\eta=-0.18$)
and confidence level contours obtained
when adding the Silver sample to the Gold sample.
}
  \label{fig:fit1}
\end{figure}

Figure~\ref{fig:fit1} shows the result of fitting $\sigma_*$ and $\eta$ to the Gold sample. The best-fitting model, with  
$\chi^2_{\rm min}=23.71$
for 
22 degrees of freedom corresponding to a reduced $\chi^2$ of 
1.08
is indicated by the circle. 
The solid contours show 68.3, 95, and 99 per cent confidence levels 
corresponding to $\chi^2=\chi^2_{\rm min}+\Delta \chi^2$ with $\Delta \chi^2=2.30, 5.99,$ and 9.21, respectively. 
The best-fitting values are:
$\sigma_*=136$ km s$^{-1}$ and $\eta=0.27$. 
We find $\sigma_* \la 190$ km s$^{-1}$ at the 95 per cent confidence level. 
Since $\kappa \propto \sigma_*^2$, a vanishing value  of $\sigma_*$ corresponds to no gravitational lensing in our model. At the 95 per cent confidence level, the results are therefore consistent with no gravitational lensing.

\begin{figure}
\includegraphics[width=84mm,angle=-90]{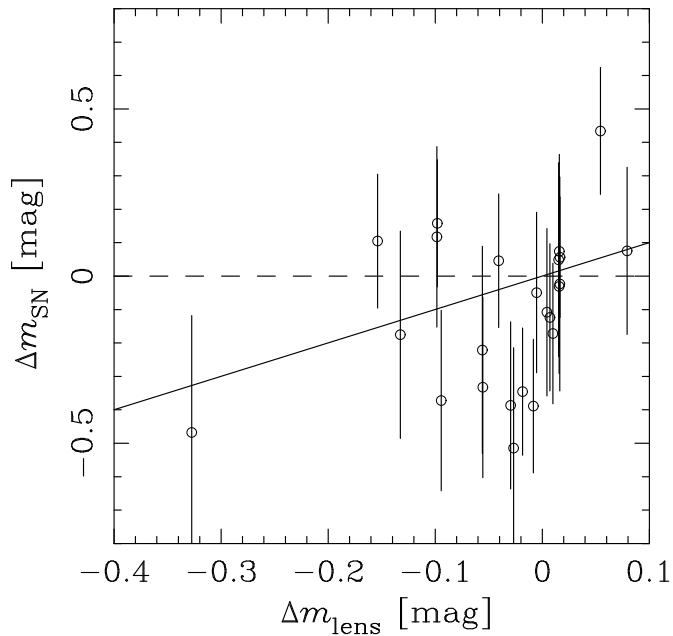}                                               
 \caption{
 Magnification-residual diagram for the best-fitting halo model to the Gold sample. The solid and dashed line have slope unity and zero, respectively.
 }
  \label{fig:magres}
\end{figure}

If the lensing hypothesis is correct, residuals and magnifications should be correlated. Figure~\ref{fig:magres} shows a magnification-residual diagram computed for the best-fitting halo model for the Gold sample.  We expect the points to scatter around the solid line, if they are correlated. 
The linear correlation coefficient is $r_{\rm corr}=0.36$ for this sample, which is similar to what 
\citet{jon07} found assuming a halo model based on Tully--Fisher \citep{tul77} and Faber--Jackson 
\citep{fab76} relations. For simulated data sets corresponding to no gravitational lensing we find
$r_{\rm corr} \ge 0.36$ for the best-fitting model for 18 per cent of the samples. 

Our estimates of $\sigma_*$ and $\eta$ can be translated into a relation between luminosity and mass.
The (aperture) mass enclosed within a radius $R$ is for a SIS given by $M(r \le R)=2\sigma^2R/G$. For the scaling model described by 
equation~(\ref{eq:sislaw}), the aperture mass can consequently be written
\begin{equation}
M(r \le R)=M_*\left(\frac{L}{L_*}\right)^{2\eta},
\end{equation}
where $M_*=2\sigma_*^2R/G$ is the mass of a $L_*$ galaxy halo. For the Gold sample we find  
$M_*=1.3\times10^{12}h^{-1}M_{\sun}$ ($M_*\la2.5 \times10^{12}h^{-1}M_{\sun}$ at the 95 per cent confidence level)
assuming $R=150h^{-1}$ kpc.

In order to increase the number of SNe~Ia we could include also SNe~Ia classified as silver by \citet{rie07} in our fit. 
The total number of silver SNe~Ia in the GOODS fields is 10, but 4 of these are located to close to the boundary of the field. 
The residual of HST05Red is $\Delta m_{\rm lens} < -1$ mag. 
Since there are no massive galaxies along the line of sight, 
such a large magnification is very unlikely to be caused by lensing.
Consequently, we do not include HST05Red in the analysis.
Our Silver sample therefore contains 5 objects.
The best-fitting values to the joint Silver and Gold sample is indicated by the star in Fig.~\ref{fig:fit1}. The 68.3, 95, and 99 per cent confidence levels are represented by the dotted contours. 
Including the 5 objects in the Silver sample have a very small effect on the confidence level contours, but leads to shifted best-fitting values ($\sigma_*=73$ km s$^{-1}$ and $\eta=-0.18$).

% SIS figure
%%%%%%%%%%%%%%%%%%%%%%%%%%%%%%%%%%%%%%%%%%%%%%%%%%%%%%%%%%%%%%%%%%%%%%%%%%%%%

% Magnification-residual diagram
%%%%%%%%%%%%%%%%%%%%%%%%%%%%%%%%%%%%%%%%%%%%%%%%%%%%%%%%%%%%%%%%%%%%%%%%%%%%%

%%%%%%%%%%%%%%%%%%%%%%%%%%%%%%%%%%%%%%%%%%%%%%%%%%%%%%%%%%%%%%%%%%%%%%%%%%%%%
%%%%%%%%%%%%%%%%%%%%%%%%%%%%%%%%%%%%%%%%%%%%%%%%%%%%%%%%%%%%%%%%%%%%%%%%%%%%%
\section{Systematic effects} \label{sec:sys}
In this section we examine different systematic uncertainties which could potentially affect or results. 
Since light curve fitting is a particularly important source of systematic uncertainty, we defer the discussion of it to section~\ref{sec:union}.

\subsection{Magnification factor normalisation}\label{sec:norm} 
  To compute the magnification normalisation, $\langle \mu\rangle$, 
we use simulated lines of sight. 
The simulated lines of sight are randomly located in the image plane, but should be randomly located in the source plane. This leads to an over estimation of $\langle \mu \rangle$, because
magnified lines of sight subtend a larger fraction of the image plane
than de-magnified ones. To compensate for this a weighting procedure
can be used. For halo models predicting large magnification factors,
the weighting procedure rejects most of the lines of sight. The
convergence of $\langle \mu \rangle$ is consequently very slow for
these models. Therefore we use the weak lensing approximation, without
any weighting, instead. 
We have checked that this approach is reliable.

\begin{figure}
\includegraphics[width=84mm,angle=-90]{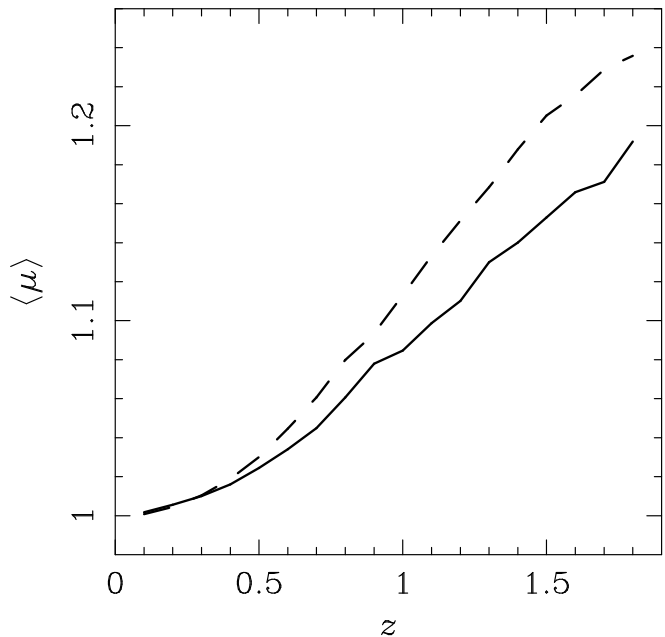}                                                
 \caption{Average 
 magnification factor, $\langle \mu \rangle$,  
 computed for simulated lines of sight.
 Solid and dashed curves correspond to GOODS--S and GOODS--N, respectively.  
The curves correspond to a halo model with parameters
 $\sigma_*=136$ km s$^{-1}$ and $\eta=0.27$.
 }
  \label{fig:bg}
\end{figure}

Figure~\ref{fig:bg} shows 
 $\langle \mu\rangle$
as a function of redshift for GOODS--S (solid curve) and GOODS--N (dashed curve) computed
for the best-fitting model to the Gold sample ($\sigma_*=136$ km s$^{-1}$ and $\eta=0.27$).
For the two fields the  normalisation
clearly differs. The normalisation
depends on the magnitude limit of the survey and cosmic variance.  
For this model the weak lensing approximation leads to under estimated values of 
$\langle \mu \rangle$ by less than 1 and 2 per cent at $z_{\rm SN}=1$ and $z_{\rm SN}=1.5$, respectively. The error from the approximation increases for halo models predicting larger magnifications. For example, the errors for a model with $\sigma_*=200$ km s$^{-1}$ and $\eta=0.27$ 
are 3 and 6 per cent for $z_{\rm SN}=1$  and $z_{\rm SN}=1.5$,  respectively. 

To compute 
$\langle \mu \rangle$
for a particular model and a particular redshift, 1000 simulated 
lines of sight are used. Since increasing this number by a factor of 2 has negligible impact on the results, we 
conclude that 1000 simulated lines of sight are sufficient for
estimating 
$\langle \mu \rangle$.

% Background term figure
%%%%%%%%%%%%%%%%%%%%%%%%%%%%%%%%%%%%%%%%%%%%%%%%%%%%%%%%%%%%%%%%%%%%%%%%%%%%%

\subsection{Cosmology}\label{sec:cosmo}
Cosmological parameters enters our calculations not only through the angular diameter distances, but also, more importantly,
through the Hubble diagram residuals. Ideally the residuals would be computed using the average SN~Ia brightness, but the 
size of the data set we use is too small to permit this approach. Instead we have to assume a cosmological model in our calculations.
Since we restrict our investigation to a flat universe with a cosmological constant, the cosmological model
can be simply characterised by $\Omega_{\rm M}$. 
Changing the value of $\Omega_{\rm M}$ from 0.3, which is what we usually assume, to 0.25 and 0.35
results in a change in the best-fitting value of $\sigma_*$ from 136 to 152 and 129 km s$^{-1}$, respectively. 
The corresponding changes in the best-fitting values of $\eta$ are negligible.     
The cosmological model thus has a rather weak effect on the results. For the Gold sample, the
effect of the cosmological model when considering realistic deviations from the concordance model, is certainly smaller than the statistical uncertainty.

\subsection{Cut-off radius}
When deciding which galaxies to include in the analysis, we use an angular cut-off radius 
$\theta_{\rm c}$. If a circle with this radius centred at the SN~Ia position crosses the edge of the field or a masked region, the SN~Ia is excluded from the analysis. Since the SIS profile is divergent, we also use a  cut-off radius, $\xi_{\rm max}$, in the lens plane. At low redshifts $\xi_{\rm max}$ has no effect since its corresponding angular scale is larger than $\theta_{\rm c}$. 
We use $\xi_{\rm max}=150h^{-1}$ kpc. Increasing the value of $\xi_{\rm max}$ to $250h^{-1}$ or $350h^{-1}$ kpc has a negligible effect on the results.

\subsection{Magnitude limit}
Our galaxy catalogues have a limiting magnitude of 25 mag in the $K_{\rm S}$-band. 
We assume that the more luminous a galaxy is, the more massive and more important it is for the gravitational lensing calculations.
To test this assumption we can compare results obtained using different magnitude limits. 
Table \ref{tab:ks} shows the best-fitting values of $\sigma_*$ and $\eta$ for magnitude limits in the range
$23.5 \le K_{\rm S} \le 25$.  The table also shows the number of early and late type galaxies in the GOODS fields brighter than the magnitude limit. 
The value of $\eta$ increases slightly when the limiting magnitude increases, while $\sigma_*$ decreases. The gravitational lensing effect from galaxies too faint to be included in the calculation appears to be compensated for by increased values of $\sigma_*$.
Since the best-fitting values of the halo parameters approach the ones obtained for $K_{\rm S}=25$ mag when the limiting magnitude increases 
and the scatter in the best-fitting values is much smaller
than the statistical uncertainty, we conclude that this magnitude limit is sufficient for our study. 
From the table it is clear the number of early type galaxies is less affected by the magnitude limit than the number of late type galaxies. 

% K_S  table
%%%%%%%%%%%%%%%%%%%%%%%%%%%%%%%%%%%%%%%%%%%%%%%%%%%%%%%%%%%%%%%%%%%%%%%%%%%%%
\begin{table}
 \caption{
 Effect of changing the limiting magnitude. The table shows the best-fitting values of 
 $\sigma_*$ and $\eta$
 obtained for different magnitude limits in the $K_{\rm S}$-band. The number of early ($N_{\rm early}$) and late ($N_{\rm late}$) type galaxies in the GOODS fields brighter than the limit are also given in the table.
 }
 \label{tab:ks}
 \begin{tabular}{@{}ccccc}
  \hline
    $K_{\rm S}$ (mag) & $\sigma_*$ (km s$^{-1}$)& $\eta$ & $N_{\rm early}$ & $N_{\rm late}$ \\
  \hline
23.5 &152 & 0.21 & 634 & 2296 \\
24.0 &149 & 0.24 & 752 & 3638 \\
24.5 &136 & 0.30 &  825 & 5659 \\
25.0 &136 & 0.27 & 883 & 8502 \\
  \hline
 \end{tabular}
 \end{table}

\subsection{Photometric redshift uncertainties}
\begin{figure}
\includegraphics[width=84mm,angle=-90]{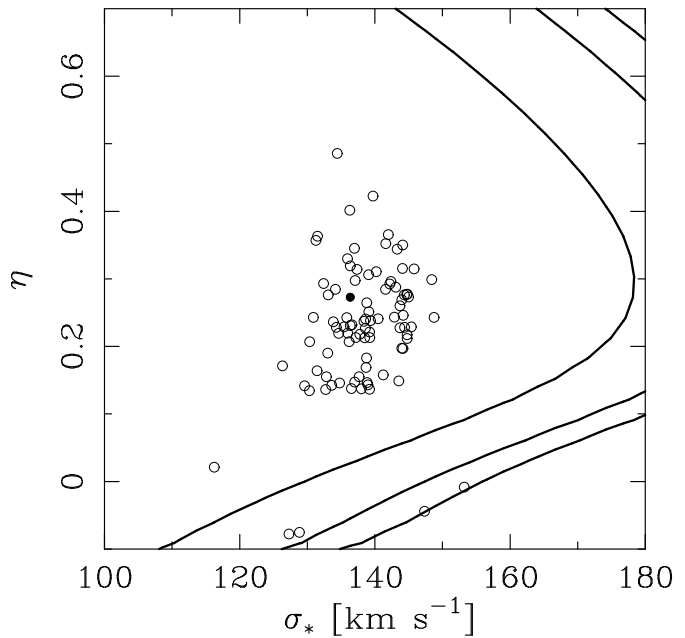}                                                
 \caption{Best-fitting parameters $\sigma_*$ and $\eta$ computed for different realisations of the galaxy catalogues
 according to the photometric redshift probability distributions of the galaxies. Note that the scale is different from Fig.~\ref{fig:fit1}.
 }
  \label{fig:photz}
\end{figure}

The redshift of the galaxies is a particularly important parameter in the calculation of 
the magnification,
because it enters the calculations
not only via the angular diameter distances involved, but also through the estimation of the luminosity of the galaxies. 
For most of the galaxies we have to use  photometric redshifts with considerable uncertainty.  
The error in the SN~Ia
redshift is negligible. 
To investigate the effect of the photometric redshifts on the dark matter constraints, we use simulated galaxy catalogues where
the original photometric redshifts, $z_{\rm gal}$, have been replaced by redshifts, $z_{\rm gal}'$, drawn at random from the probability distribution of the photometric redshifts. A change in the redshift is accompanied by a correction, $\Delta M_B(z_{\rm gal}')$, in the absolute magnitude. The absolute
magnitude of the galaxy with the new redshift is hence 
\begin{equation}
M_B(z_{\rm gal}')=M_B(z_{\rm gal})+\Delta M_B(z_{\rm gal}').
\end{equation}
The results of the simulations are presented in Fig.~\ref{fig:photz}. Open circles represents the best-fitting values of $\sigma_*$ and $\eta$ 
to 100 
simulated galaxy catalogues. For the benefit of comparison, confidence level contours and best-fitting values (solid circle) obtained for  the original catalogue are also
shown in the figure. From 
Fig.~\ref{fig:photz} we conclude that the uncertainty in the results due to the photometric redshifts is small compared to the statistical uncertainty.

% Photo-z figure
%%%%%%%%%%%%%%%%%%%%%%%%%%%%%%%%%%%%%%%%%%%%%%%%%%%%%%%%%%%%%%%%%%%%%%%%%%%%%

\subsection{Shifted halo and galaxy positions}
Shifts between galaxy and halo positions could potentially affect or results. 
In a study of strong gravitational lensing systems, 
\citet{yoo06} found the centre of light to coincide with the centre of mass. Even
if galaxies and haloes are not shifted relative to each other, measurement errors could lead to a shift. We have investigated 
the effect of shifting all galaxy positions by a value drawn from a Gaussian distribution with a standard deviation of $0.5\arcsec$.
We find the effect of shifted galaxy positions to be negligible.

\subsection{Ellipticity}
So far we have assumed haloes to be axially symmetric. In order to test for the effect of deviations from axial symmetry, we 
make use of a generalisation of the SIS called the singular isothermal ellipsoid \citep[SIE,][]{kor94}. For the SIE model    
the convergence,
\begin{equation}
\kappa_{\rm gal}(x,\phi)=\frac{1}{2x}\frac{\sqrt{f}}{\sqrt{\cos^2\phi+f^2\sin^2\phi}},
\label{eq:sie}
\end{equation}
is a function 
not only of $x=\xi/\xi_0$, with $\xi_0$ given by equation~(\ref{eq:sisxi}), but also of the angle, $\phi$, between
 the position of the ray in the lens plane
and the minor axis of the lens. In addition to $\sigma$, the SIE model depends on the minor to major axis 
ratio, $f$. 
The shear components of the SIE model are given by 
$\gamma_{{\rm gal},1}(x,\phi)=-\kappa_{\rm gal}(x,\phi) \cos(2\phi)$
 and  $\gamma_{{\rm gal},2}(x,\phi)=-\kappa_{\rm gal}(x,\phi) \sin(2\phi)$.

We 
use simulations to explore the effects of elliptical haloes. 
For simplicity we assume all values of  the inclination of the galaxies, $\psi$, in the range 
$0 < \psi \le 2\pi$ to be equally plausible. Numerical simulations \citep[e.g.,][]{gus06} implies that dark matter haloes 
are triaxial. \citet{gus06} found $c/a=0.73 \pm 0.11$, where $c$ and $a$ are the smallest and largest of the principal axis
of a triaxial halo. We make the conservative assumption that we are viewing all halos along a line perpendicular to the plane spanned
by $a$ and $c$, hence maximising the effect of the ellipticity.  In our simulations, the axis ratio is therefore given by a Gaussian distribution with mean 0.73 and standard deviation 0.11 
truncated at $f=1$. For this choice of parameters, the convergence of the SIE is on average 7 per cent smaller than a SIS with the same velocity dispersion. 

\begin{figure}
\includegraphics[width=84mm,angle=-90]{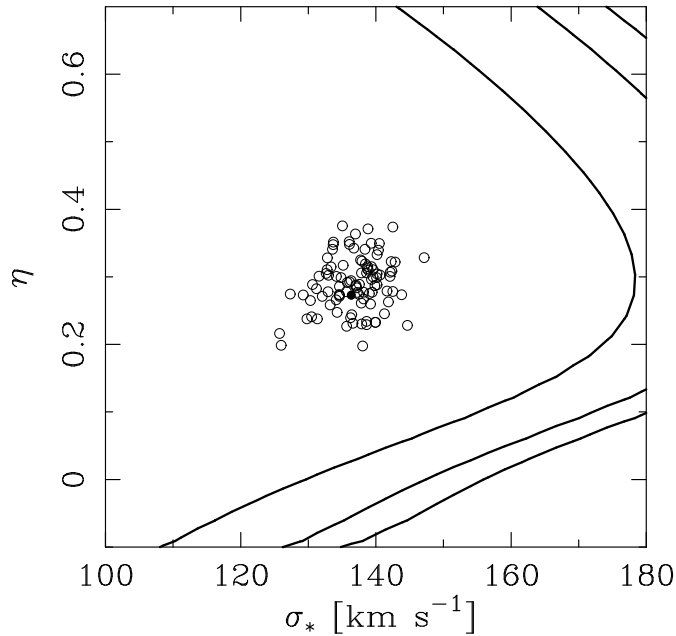}                                                
 \caption{Best-fitting parameters $\sigma_*$ and $\eta$ to the SIE model computed for different realisations of the galaxy catalogues
 with simulated values of $f$ and $\psi$. Note that the scale is different from Fig.~\ref{fig:fit1}.
 }
  \label{fig:sie}
\end{figure}

The best-fitting parameters (indicated by open circles) to  100 simulated data sets are shown in 
Fig.~\ref{fig:sie}. The points corresponding to
the simulated SIE models are scattered around the solid circle, which indicates the best-fitting parameters to the SIS model. 
Since the scatter is small compared to the statistical uncertainty, 
we conclude that the ellipticity of the galaxies can safely be neglected.

Although the ellipticity of the haloes is not a concern for the current data set, it could certainly be for larger data sets with smaller statistical uncertainty. In that case the observations of galaxy ellipticities, which are used for shear measurements, might be used to
improve the modeling of the galaxies for the purpose of convergence measurements. 

% SIE figure
%%%%%%%%%%%%%%%%%%%%%%%%%%%%%%%%%%%%%%%%%%%%%%%%%%%%%%%%%%%%%%%%%%%%%%%%%%%%%

\subsection{Extinction by dust}\label{sec:dust}
Another potential systematic uncertainty is dimming of SNe~Ia by dust
extinction \citep{gob02,nor08}. Such
extinction may take place in the host galaxy, in the Milky Way, in
intervening galaxies and in intergalactic dust
\citep{mor03,ost05}. The distance moduli of
the SNe~Ia included in the analysis are corrected for dust extinction
in the host galaxy based on the colours of the SN~Ia and for dust in
the Milky Way. Since the host galaxy dust corrections are based on the
observed colours of the SNe~Ia, they will also to some extent correct
for dust in intervening galaxies and intergalactic dust. However,
since we do not expect this correction to be exact, SNe~Ia with
foreground galaxies close to the line of sight may suffer from
systematic effects from the dust extinction in the intervening
galaxies. This could lead to a correlation between the residual
extinction and magnification.

Most SNe~Ia in the Gold sample have an impact parameter of the closest
galaxy along the line of sight that is larger than the disk length
scale \citep[$\la 5 h^{-1}$ kpc,][]{fre70}. We therefore expect the
extinction from dust in the disks of intervening galaxies to have a
small effect for most SNe in the current sample.  For SN2003es and
HST05Lan, the closest impact parameter is $5h^{-1}$ and $7h^{-1}$ kpc,
respectively. Since these values are comparable to the disk length
scale, SN2003es and HST05Lan could be affected by disk dust extinction.
Both SNe~Ia are $\sim 0.1$ mag dimmer than expected from the concordance model.
According to the best-fitting halo model SN2003es and HST05Lan should be
magnified by $-0.10$ and $-0.15$ mag, respectively.
 For SN2003es the
closest galaxy is responsible for roughly  $2/3$ of the magnification.
The contribution from the closest galaxy to the magnification of
HST05Lan is negligible.
For these two SNe~Ia there is hence a large discrepancy between their measured brightness and predicted magnification. 
One possible explanation of this discrepancy is
that  the reddening occurs at a redshift different from that of the host galaxy.
If these two SNe~Ia are removed from the Gold sample, the best-fitting
values of $\sigma_*$ and $\eta$ become
$154$ km s$^{-1}$ and $-0.06$, respectively.

Since $30 < \sigma_* <220$ km s$^{-1}$
at the 99 per cent confidence level, this reduced sample is strongly
supporting the gravitational lensing hypothesis. 
The
quality of the fit is significantly improved when removing SN2003es
and HST05Lan from the Gold sample, more so than for the removal of any 
other two
SNe in the sample. 

\citet{men09a} recently reported the detection of dust in the halos of 
galaxies,
at distances from 20 kpc to several Mpc from galaxy centres, based on
cross-correlating the brightness of high redshift quasars with the
position of foreground galaxies \citep[see also][]{ost06,ost08}. The dust extinction
was found to have a very similar dependence on the impact parameter as
the lensing magnification. However, since the dust extinction is an
order of magnitude smaller than the magnification, the effect for SNe
with large impact parameters will be negligible in this study.

The implications for SN~Ia cosmology due to this detection was 
investigated in
\citet{men09b}, where it was found that the effects of the intergalactic 
dust
may not be properly accounted for by the SN~Ia colour corrections,
leading to biased cosmological parameters. Such a bias could be
relevant for the study presented here, since it would affect the
Hubble diagram residuals. \citet{men09b} found the bias in the
cosmological parameters due to dust extinction to be a few
percent. According to 
 section~\ref{sec:cosmo}
shifts in $\Omega_{\rm M}$
of that order give only small changes in the results.

%%%%%%%%%%%%%%%%%%%%%%%%%%%%%%%%%%%%%%%%%%%%%%%%%%%%%%%%%%%%%%%%%%%%%%%%%%%%%
%%%%%%%%%%%%%%%%%%%%%%%%%%%%%%%%%%%%%%%%%%%%%%%%%%%%%%%%%%%%%%%%%%%%%%%%%%%%%
\section{Light curve fitting} \label{sec:union}
Light curve fitting is an important ingredient in supernova cosmology. Since light curve fitting affects the Hubble diagram residuals, it is also a potential systematic uncertainty for the analysis presented here. In the previous analysis we have used distance moduli from \citet{rie07}, which were obtained using the MLCS2k2 \citep{jha07} light curve fitting package. Recently \citet{kow08} presented a compilation of SNe~Ia, which includes the GOODS SNe~Ia, called the Union data set.  For the Union data set a different light curve fitting package, named SALT \citep{guy05}, was used.
In the following we compare results on dark matter halo properties obtained using different light curve fitting packages
[see \citet{kes09} for a thorough study of the systematic differences between them].

Since \citet{rie07} and \citet{kow08} use different selection criteria, different GOODS SNe~Ia are included in the Gold and Union samples. 
For example, SN2003be, SN2003bd, SN2003es, and HST04Tha were rejected by \citet{kow08}, because their light curves had insufficient early data, but included by \citet{rie07} in the Gold sample.
A column indicating which GOODS SNe~Ia belong to which sample is included 
In table~\ref{tab:sn}. HST04Sas is located too close to a masked region and is for that reason excluded from the comparison below. 

\begin{figure}
\includegraphics[width=84mm,angle=-90]{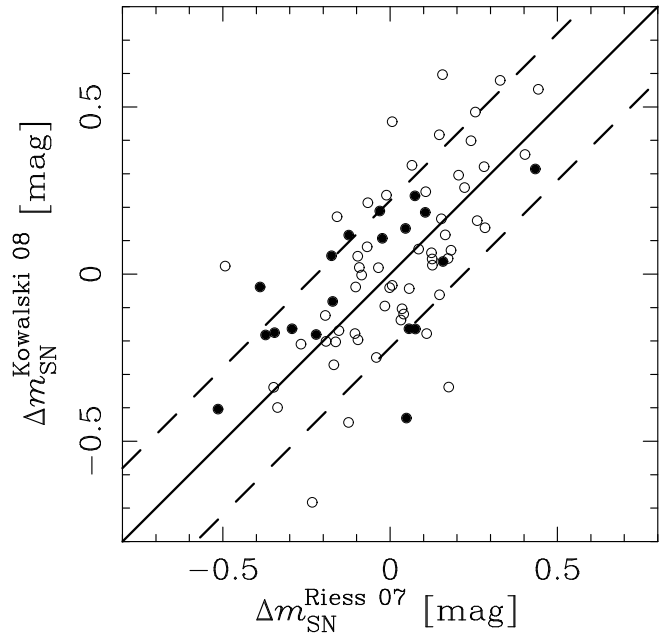}                                                
 \caption{
 Scatter plot of SN~Ia Hubble diagram residuals from \citet{kow08} versus residuals from \citet{rie07}. Circles represent the 76 SNe~Ia common to the Gold and Union data sets.
 Filled circles correspond to the subset of 19 SNe~Ia (including HST04Sas) located in the GOODS fields. The dashed lines outlines the 
 scatter ($0.22$ mag) between the data sets.
 }
  \label{fig:kow}
\end{figure}

Figure~\ref{fig:kow} shows  a scatter plot of SN~Ia Hubble diagram residuals from 
 \citet{kow08} versus residuals from \citet{rie07} for the 
 76 SNe~Ia common to the Gold and Union data sets. 
 We expect the points to scatter around the solid line (with slope unity), if there is no systematic offset between the two sets of residuals. The \citet{kow08} residuals are on average
 $0.013$ mag dimmer than the \citet{rie07} residuals. The scatter around the solid line, which is indicated by the dashed lines, is $0.22$ mag. For the subset of SNe~Ia located in the GOODS fields, indicated by filled circles, there appears to be a larger offset. The \citet{kow08} residuals are on average $0.055$ mag dimmer for this subset of SNe~Ia. Since the scatter is quite large, 
 $0.20$ mag, this offset might not be significant.

% Kowalski-Riess figure
%%%%%%%%%%%%%%%%%%%%%%%%%%%%%%%%%%%%%%%%%%%%%%%%%%%%%%%%%%%%%%%%%%%%%%%%%%%%%

\begin{figure}
\includegraphics[width=84mm,angle=-90]{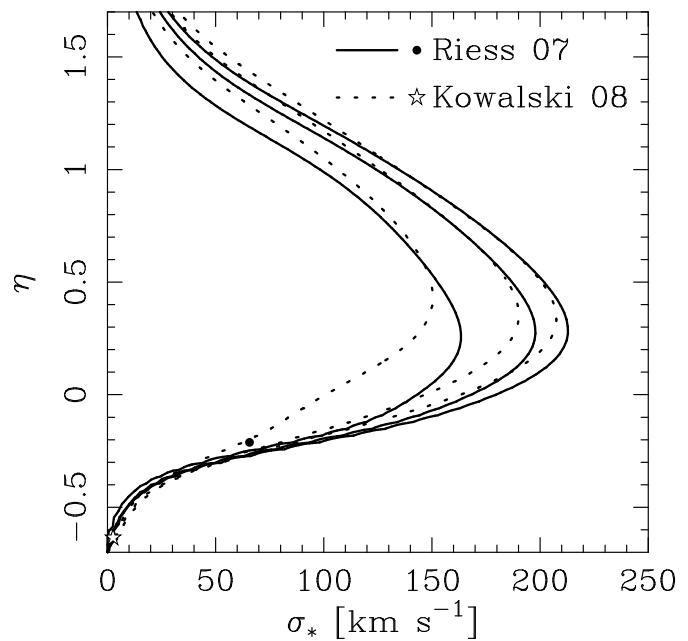}                                                
 \caption{Comparison of results obtained for 18 SNe~Ia with different residuals.
 The circle and the contours show the best-fitting values and the 68.3, 95, and 99 per cent
confidence level for residuals from \citet{rie07}. 
The star and the dotted contours show the best-fitting 
values and the 68.3, 95, and 99 per cent confidence level for residuals from \citet{kow08}. 
 }
  \label{fig:joint}
\end{figure}

Figure~\ref{fig:joint} shows a comparison between results obtained using residuals from \citet{rie07} and \citet{kow08} for the 18 GOODS SNe~Ia
in common to both data sets.
The circle and the star correspond to the best-fitting values obtained for residuals from
\citet{rie07} and \citet{kow08}, respectively.
Solid and dotted contours show 68.3, 95, and 99 per cent confidence level
contours corresponding to the data from \citet{rie07} and \citet{kow08}.
The different residuals lead to somewhat different results,

% Joint figure
%%%%%%%%%%%%%%%%%%%%%%%%%%%%%%%%%%%%%%%%%%%%%%%%%%%%%%%%%%%%%%%%%%%%%%%%%%%%%

\begin{figure}
\includegraphics[width=84mm,angle=-90]{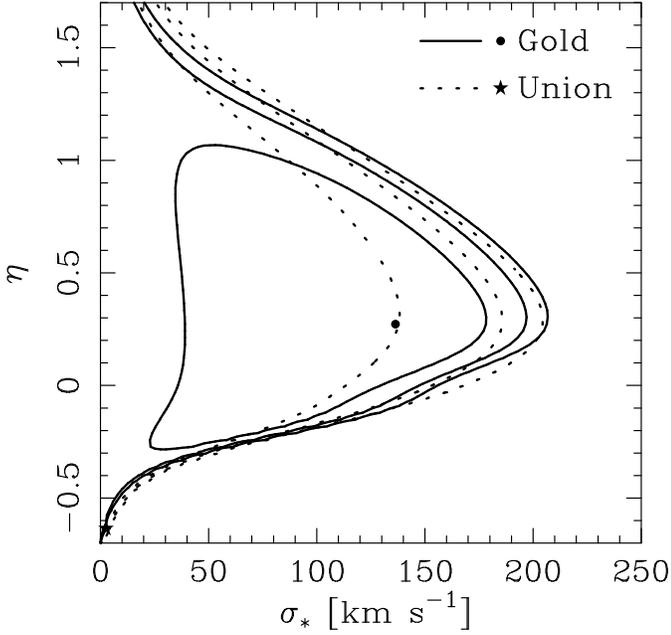}                                                
 \caption{Comparison between the results obtained from the Gold and the Union sample.
The circle and the contours show the best-fitting values and the 68.3, 95, and 99 per cent
confidence level for the Gold sample. The star and the dotted contours show the best-fitting 
values and the 68.3, 95, and 99 per cent confidence level for the Union sample. 
 }
  \label{fig:kowfit1}
\end{figure}

Figure~\ref{fig:kowfit1} shows a comparison between the results obtained for the Gold (circle and solid contours) and the Union (star and dotted contours) samples. 
The best-fitting value for the Union sample is 
$\sigma_*=3$ km s$^{-1}$ and $\eta=-0.64$. 
The results obtained using different light curve fitting packages clearly disagree at
 the 68.3 per cent confidence level.

 Adding an offset of $-0.055$ mag, as suggested by Fig.~\ref{fig:kow}, to the \citet{kow08} residuals has little effect on the best-fitting values of the halo parameters, but improves the agreement between the Gold and Union results at the
68.3 per cent confidence level.

% Kowalski Fit1
%%%%%%%%%%%%%%%%%%%%%%%%%%%%%%%%%%%%%%%%%%%%%%%%%%%%%%%%%%%%%%%%%%%%%%%%%%%%%

\begin{figure}
\includegraphics[width=84mm,angle=-90]{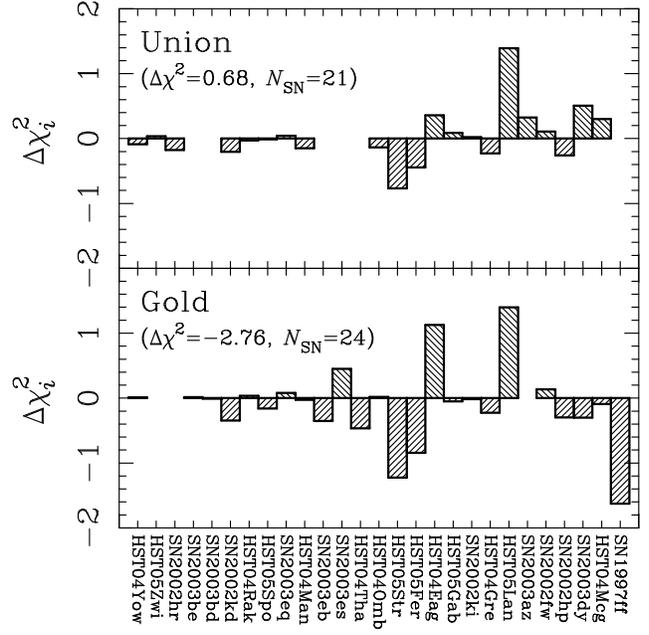}                                                
 \caption{Contributions from individual SNe~Ia to 
 $\Delta \chi^2=\chi^2(\blambda_{\rm halo}^{\rm bf})-\chi^2(\blambda_{\rm halo}^{\rm null})$ for the Union (top panel) and Gold (bottom panel) samples. The best-fitting model, 
 $\blambda_{\rm halo}^{\rm bf}$, to the Gold sample was used to compute the differences, 
 $\Delta \chi^2_i$.
 }
  \label{fig:chi2bf}
\end{figure}

The Gold and Union samples consist of 24 and  21 
SNe~Ia, respectively. Since they have only 18 SNe~Ia in common, the Gold sample contain 6 objects which are not in the Union sample, while 
 3
 objects are unique to the Union sample. Since these data sets are very small we would like to investigate the effects of individual SNe~Ia in more detail. In order to do this we study the difference in 
$\chi^2$ for the best-fitting model to the Gold sample 
and the no lensing,  or null, hypothesis, 
$\Delta \chi^2=\chi^2(\blambda_{\rm halo}^{\rm bf})-\chi^2(\blambda_{\rm halo}^{\rm null})$. If 
$\Delta \chi^2<0$ the best-fitting model is favoured, whereas the null hypothesis is preferred by the data  if $\Delta \chi^2 >0$. Since $\Delta \chi^2$ is a sum, we can study the 
contribution, $\Delta \chi^2_i$, to it from individual SNe~Ia. The diagram in Fig.~\ref{fig:chi2bf} shows  $\Delta \chi^2_i$ for the Union and Gold sample in the top and bottom panel, respectively. For both samples $\Delta \chi^2_i$ was computed using the best-fitting values
of the halo parameters
 to the Gold sample. The gaps in the diagram represents the SNe~Ia that were not included in either the Union or Gold sample.
This diagram reveals some important differences between the two data sets. For the Gold sample  
$\Delta \chi^2=-2.81$ 
with HST05Str,  HST05Fer, and SN1997ff giving large negative contributions. HST05Lan 
and HST04Eag, on the other hand, give large positive contributions.  
For the Union sample only HST05Str  and HST05Fer give large negative contributions, because SN1997ff is not included in this sample.  
HST05Lan gives a large positive contribution to $\Delta \chi^2$ also for the Union sample. 
It is clear from the diagram that SN1997ff is responsible for much of the difference in $\Delta \chi^2$
between the two samples.
The results thus depends
on the contributions from individual SNe~Ia, to a relatively large extent.    

\begin{figure}
\includegraphics[width=84mm,angle=-90]{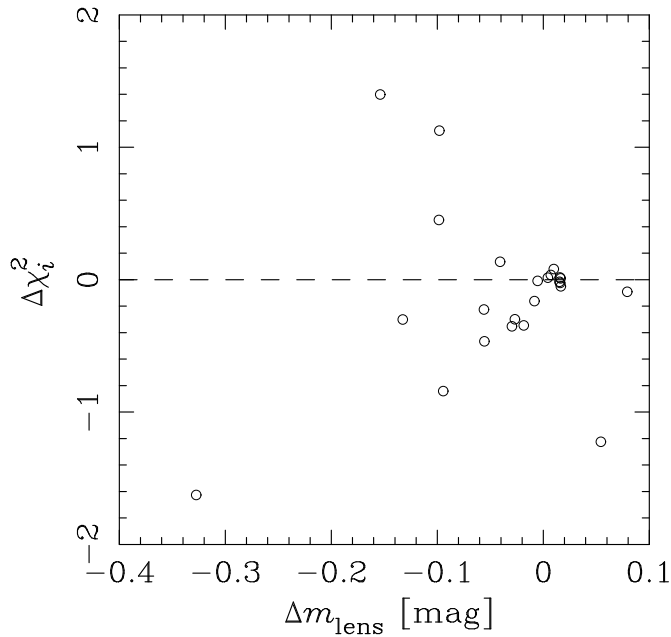}                                                
 \caption{ 
 Scatter plot of contributions from individual SNe~Ia to 
 $\Delta \chi^2=\chi^2(\blambda_{\rm halo}^{\rm bf})-\chi^2(\blambda_{\rm halo}^{\rm null})$  versus magnification computed for the best-fitting model for the Gold sample. 
 }
  \label{fig:chi2mag}
\end{figure}

Figure~\ref{fig:chi2mag} shows a scatter plot of $\Delta \chi^2_i$ versus $\Delta m_{\rm lens}$ for the 
Gold sample. Since there is no sign of any correlation between $\Delta \chi^2_i$ and $\Delta m_{\rm lens}$, SNe~Ia contribute both positively and negatively to $\Delta \chi^2$ regardless of whether they are magnified or de-magnified.

In section~\ref{sec:dust} we discussed the possibility of HST05Lan  being affected by dust extinction. Extinction by dust could be the reason why 
HST05Lan opposes the gravitational lensing hypothesis. 
Excluding this SN~Ia from the Union data set
leads to  best-fitting 
values $\sigma_*=84$ km s$^{-1}$ and $\eta=-0.15$, which are in slightly better agreement with the gravitational lensing 
hypothesis.

% Chi2 bf  figure
%%%%%%%%%%%%%%%%%%%%%%%%%%%%%%%%%%%%%%%%%%%%%%%%%%%%%%%%%%%%%%%%%%%%%%%%%%%%%

% Gold chi2 vs magnification figure
%%%%%%%%%%%%%%%%%%%%%%%%%%%%%%%%%%%%%%%%%%%%%%%%%%%%%%%%%%%%%%%%%%%%%%%%%%%%%

%%%%%%%%%%%%%%%%%%%%%%%%%%%%%%%%%%%%%%%%%%%%%%%%%%%%%%%%%%%%%%%%%%%%%%%%%%%%%
%%%%%%%%%%%%%%%%%%%%%%%%%%%%%%%%%%%%%%%%%%%%%%%%%%%%%%%%%%%%%%%%%%%%%%%%%%%%%
\section{Summary and Discussion} \label{sec:disc}
Distant SNe~Ia at high redshift might be affected by gravitational lensing due to matter in the foreground. Gravitational lensing magnification or de-magnification should correlate with Hubble diagram residuals. By exploiting this correlation \citep{jon07}, we measure the relation between galaxy luminosity and the velocity dispersion of dark matter haloes using data from the GOODS.  

For a scaling law, $\sigma=\sigma_*(L/L_*)^{\eta}$, describing the relationship between galaxy 
absolute $B$-band luminosity, $L$, and dark matter velocity dispersion, 
we find the best-fitting parameters to be $\sigma_*=136$ km s$^{-1}$ and $\eta=0.27$
for $L_*=10^{10}h^{-2}L_{\sun}$. 
These results agree well with the results from galaxy--galaxy lensing studies. 
In a galaxy--galaxy lensing study \citet{kle06} found $\sigma_*=132^{+18}_{-24}$ km s$^{-1}$ and $\eta=0.37 \pm 0.15$ 
for the same scaling law and halo model as we use here. 
\citet{hoe04} also used galaxy--galaxy lensing to investigate dark matter, but used  
 the truncated isothermal sphere model
\citep{bra96}, 
which has an additional parameter describing the truncation radius, $s$, of the halo. For a value of $\eta$ fixed to 0.3 they found 
$\sigma_*=136 \pm 5 \pm 3$ km s$^{-1}$ (statistical and systematic uncertainties), which agrees with our result, and $s=185^{+30}_{-28}h^{-1}$ kpc.  

The constraints derived from galaxy--galaxy lensing are stronger than the ones derived here. Considering how small our sample of lensed SNe~Ia is, it is encouraging that we can derive any constraints at all. Our results rely upon a different aspect of gravitational lensing 
and therefore provide an independent cross-check of the galaxy--galaxy lensing results.

 We have also investigated different systematic uncertainties. The effects of the uncertainties in the cosmological model, photometric redshifts, and magnitude limit appears to be small. 
Light curve fitting is another systematic effect, which unlike the other ones cannot be ignored for the data set used here.
Different light curve fitting packages yield different distances for the same light curves. 
The Gold and Union samples, which are both small and include different GOODS SNe~Ia,
 give rather different results. The Union sample is consistent with the no lensing hypothesis ($\sigma_*=0)$ at the 68.3 per cent confidence level. 
 The Gold sample is consistent with the no lensing hypothesis at the 95, but not at the 68.3 per cent confidence level. From Fig.~\ref{fig:chi2bf} it is clear that much of the differences can be attributed to individual SNe~Ia. Clearly, light curve fitting is an issue which has to be resolved before SNe~Ia can be used to their full potential as probes of the dark side of the Universe.

\begin{figure}
\includegraphics[width=84mm,angle=-90]{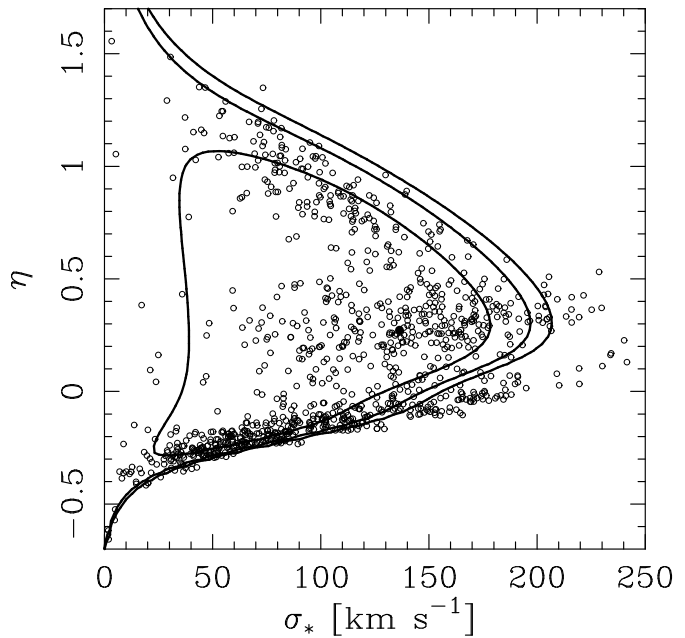}                                               
 \caption{Result of bootstrapping exercise. Open circles correspond to the best-fitting values of $\sigma_*$ and $\eta$ obtained for 1000 simulated data sets obtained by resampling of the 
 Gold sample. For comparison the best-fitting values (indicated by the filled circle) and confidence level contours for the original Gold sample are also shown.
 }
  \label{fig:boot}
\end{figure}

% Bootstrapping exercise
The effect of removing two SNe~Ia from the Gold sample 
was shown  to be quite large in section~\ref{sec:dust}.
For this reason we have performed a bootstrapping exercise to investigate the robustness of our results. 
The open circles in Fig.~\ref{fig:boot} show the best-fitting values for 1000 simulated data sets based on 
resampling of the Gold sample. To facilitate comparison, the best-fitting values (indicated by the solid circle) and confidence level contours for the original sample are also plotted in the figure. 
The highest density of points is not found near the best-fitting values of the original sample, but closer 
to $\eta \sim -0.2$ and $30 \la \sigma_*\la 120$ km s$^{-1}$. This is another indication that our results depends on individual SNe~Ia.

From our study it is clear that more high redshift SNe~Ia are needed to obtain meaningful results. 
We anticipate that the final data from the Supernova Legacy Survey \citep[SNLS,][]{ast06} will be ideally suited
for an investigation of the type presented here. According to \citet{jon08} a firm detection of the correlation between gravitational magnification and SN~Ia brightness can be expected for the final SNLS data set.
The SNLS produces not only a homogeneous set of well calibrated SNe~Ia, but also the necessary deep 
multi colour observations of the foreground galaxies.

% Boot strap
%%%%%%%%%%%%%%%%%%%%%%%%%%%%%%%%%%%%%%%%%%%%%%%%%%%%%%%%%%%%%%%%%%%%%%%%%%%%%

%%%%%%%%%%%%%%%%%%%%%%%%%%%%%%%%%%%%%%%%%%%%%%%%%%%%%%%%%%%%%%%%%%%%%%%%%%%%%
%%%%%%%%%%%%%%%%%%%%%%%%%%%%%%%%%%%%%%%%%%%%%%%%%%%%%%%%%%%%%%%%%%%%%%%%%%%%%
\section*{Acknowledgments}
AG acknowledge support from the Gustafsson foundation. AG and EM acknowledge
 financial support from the Swedish Research Council. The authors would like to thank the anonymous referee for his/her comments, which helped improving the paper.
 JJ would like to thank the Oskar Klein Centre in Stockholm for its hospitality.

%%%%%%%%%%%%%%%%%%%%%%%%%%%%%%%%%%%%%%%%%%%%%%%%%%%%%%%%%%%%%%%%%%%%%%%%%%%%%
%%%%%%%%%%%%%%%%%%%%%%%%%%%%%%%%%%%%%%%%%%%%%%%%%%%%%%%%%%%%%%%%%%%%%%%%%%%%%

\end{document}